# Patent Data for Engineering Design: A Critical Review and Future Directions

**A Preprint (Date: 2022.06.01)**

**Shuo Jiang, first author**[1]
School of Mechanical Engineering
Shanghai Jiao Tong University
800 Dongchuan Road, Shanghai, China, 200240
Data-Driven Innovation Lab
Singapore University of Technology and Design
8 Somapah Road, Singapore, 487372
shuojiangcn@gmail.com

**Serhad Sarica, second author**
Data-Driven Innovation Lab
Singapore University of Technology and Design
8 Somapah Road, Singapore, 487372
serhad_sarica@alumni.sutd.edu.sg

**Binyang Song, third author**
Department of Mechanical Engineering
Massachusetts Institute of Technology
33 Massachusetts Ave, Cambridge, MA, USA, 02139
binyangs@mit.edu

**Jie Hu, fourth author**
School of Mechanical Engineering
Shanghai Jiao Tong University
800 Dongchuan Road, Shanghai, China, 200240
hujie@sjtu.edu.cn

**Jianxi Luo, fifth author**
Data-Driven Innovation Lab
Singapore University of Technology and Design
8 Somapah Road, Singapore, 487372
luo@sutd.edu.sg

---

[1] Contact email: shuojiangcn@gmail.com

## ABSTRACT

*Patent data have long been used for engineering design research because of its large and expanding size, and widely varying massive amount of design information contained in patents. Recent advances in artificial intelligence and data science present unprecedented opportunities to develop data-driven design methods and tools, as well as advance design science, using the patent database. Herein, we survey and categorize the patent-for-design literature based on its contributions to design theories, methods, tools, and strategies, as well as the types of patent data and data-driven methods used in respective studies. Our review highlights promising future research directions in patent data-driven design research and practice.*

## KEYWORDS

*Patent, Engineering Design, Data-Driven Design, Artificial Intelligence, Data Science*

## 1 Introduction

In the engineering design field, mining patent data to develop design theory and methodology has a long history, dating back to the 1950s. Altshuller and his colleagues developed the theory of inventive problem solving (TRIZ) by manually examining thousands of patent documents [1]. Over the past few decades, advancements in artificial intelligence (AI) and data science have developed new and growing opportunities for mining and analyzing big data related to engineering design for supporting design research and practice. In contrast to human-curated design repositories [2,3], patent databases provide several advantages for data-driven engineering design.

First, patent databases are large-scale repositories that accumulate over time as inventors file patent applications for their inventions. For example, from 1963 to 2020, the United States Patent and Trademark Office (USPTO) database contained over 7.7 million granted patents. Second, patents contain massive design information on technologies, systems, or processes from all domains in both textual and visual forms, the innovation activity footprints and collaboration information of inventors and organizations in bibliometrics, and their relations to prior or future designs in the form of citations. Moreover, every patent is assigned to a domain class(es) by multiple patent examiners, making patent data ready for both supervised and unsupervised machine learning applications.

In recent years, several engineering design research groups have actively explored cutting-edge data science techniques to mine patent databases for diverse applications (hereafter referred to as **patent-for-design** studies) such as design representation, design space exploration, design prior art searching, stimuli recommendation, idea generation, and evaluation [4–10]. These patent-for-design studies relied on a broad collection of methods ranging from classic statistical analysis to network analysis, natural language processing (NLP), machine learning, deep learning, and data visualization. To the best of our knowledge, there has been no systematic review of patent-for-design research literature despite its rapid growth. Therefore, we conducted this review to elucidate the state of the art and reveal the trends. We also benchmarked the status quo with frontiers of data science to identify future research opportunities and directions.

The remainder of this paper is organized as follows. Section 2 describes literature retrieval methodology. Section 3 introduces the patent data structure and summarizes the use of patents in engineering design research field. We then present a nuanced review of all patent-for-design literature categorized by their research applications, namely, design theories, design methodologies, design tools, and design strategies in Section 4. Section 5 provides a structured and integrated analysis of the methods and algorithms used in the patent-for-design publications. Section 6 maps feasible directions for future research. Finally, Section 7 concludes the paper.

## 2 Literature Retrieval Methodology

To retrieve prior patent-for-design publications, we used the following search process to ensure comprehensiveness and relevance (retrieval date: Oct 23, 2021). Because this review study aims to contribute to the engineering design field, we started a literature search from eight leading engineering design journals: (1) ASME Journal of Mechanical Design (JMD); (2) ASME Journal of Computing and Information Science in Engineering (JCISE); (3) Research in Engineering Design (RIED); (4) Journal of Engineering Design (JED); (5) Design Studies; (6) Design Science; (7) Artificial Intelligence for Engineering Design, Analysis and Manufacturing (AI-EDAM); (8) Computer-aided design (CAD); and three conferences: (1) International Conference on Engineering Design (ICED), (2) International Design Conference (DESIGN), and (3) International Design Engineering Technical Conferences & Computers and Information in Engineering Conference (IDETC/CIE). All authors of the present study reached an agreement after discussing the journals and conferences for inclusion and conducting literature search.

To search for journal papers, in the first round, we ran a query search in Web of Science. The following query was used: *(TI="patent" OR AB="patent") AND (SO=X) AND (PY=1950-2021)*[2], where *X* represents one of eight engineering design journals. The search returned 46 articles. We manually checked these papers and identified 32 that met the scope of this review. The inclusion criteria (with an AND condition) are as follows: (1) patent data are used as research data in the body text of the paper; (2) the paper contributes to the support of engineering design-related processes. These 32 papers formed the initial core paper list.

In the second round, we browsed the journal websites to search for any missing relevant literature in the first round, including recently accepted manuscripts that are relevant to our topic but not yet included in Web of Science. We queried the search engines of the journal websites using the keyword “patent”. In this round, we found 18 additional relevant papers, bringing the list to 50.

In the third round, we removed journal restraint in the query to conduct a global search in Web of Science. The following query was used: (*TS=(”patent” AND “engineering design”)) AND (PY=1950-2021)*[3]. Selecting the document type as "article" returned 36 papers. Among them, 13 have already been included in the list from previous searches, and 6 are about using patent data as engineering design cases for education, which are beyond the scope of this review. From the remaining papers, we identified 5 relevant papers and added them to the list, resulting in a total of 55 papers.

[2] In this Web-of-Science query, TI stands for title, AB stands for abstract, SO stands for publication name, and PY stands for publication year.
[3] In this Web-of-Science query, TS stands for topic, and PY stands for publication year.

Finally, in the fourth round, we further checked the forward and backward citations of the papers in the core list using a snowballing process and found another three relevant papers to add to the list, resulting in a total of 58 journal papers.

We used a slightly different method to search conference papers. Because the index of these conference proceedings in Web of Science was rather ambiguous, in the first round, we directly queried “patent” as the keyword in the titles and abstracts of the papers using the search engines in conference websites. The searches returned 22, 2, and 73 papers in the ICED, DESIGN, and IDETC/CIE proceedings, respectively. Using the same inclusion criteria that we used for journal papers, we reduced the set to 12, 2, and 18 papers, totaling 32 conference papers. We further checked whether we had already identified and collected the journal versions of these papers. Consequently, we removed 11 papers from our set, reducing the list to 20 conference papers. Finally, we checked the forward and backward citations of these papers. Although we found some relevant papers in this step, these papers have already been identified in previous searches for journal papers. Therefore, we identified 20 conference papers related to the focus of this review.

In summary, we curated a patent-for-design literature list of 78 papers (including 58 journal papers and 20 conference papers) for review[4]. Table 1 summarizes the literature search process.

[4] The entire list of patent-for-design literature is presented in Fig. 4 (Section 4).

**Table 1**: Entire literature search process

| The Operation of Each Step | | | Number of Retrieved Papers (Accumulated) |
|---|---|---|---|
| **Journal** | Step 1 | Search in Web of Science and then manually identify relevant literature. *Query:(TI="patent" OR AB="patent") AND (SO=X) AND (PY=1950-2021)* | 32 |
| | Step 2 | Search in each journal's website (using "patent" as keyword) to search for possible missing literature. | 50 |
| | Step 3 | Search in Web of Science and then manually identify relevant literature. *Query:* (*TS=("patent" AND "engineering design")) AND (PY=1950-2021)* | 55 |
| | Step 4 | Manually check the forward and backward citations of the papers in the core list using a snowballing process | 58 |
| **Conf.** | Step 5 | Search in each conference's website (using "patent" as keyword), and remove the non-relevant ones and the conference-version ones of previously retrieved journal papers. | 78 |

Fig. 1 reports the distribution of these papers: A) publication years and B) journals and conferences. Fig. 2 shows the co-occurrence of keywords in all the retrieved literature. To create the keyword co-occurrence figure, we conducted lemmatization to reduce inflectional forms of similar keywords, and then removed all keywords that appeared only once in the literature. In addition, we removed the keyword "patents" because all papers had it. The co-occurrence of keywords forms a map to illustrate the leading research topics in patent-for-design literature.

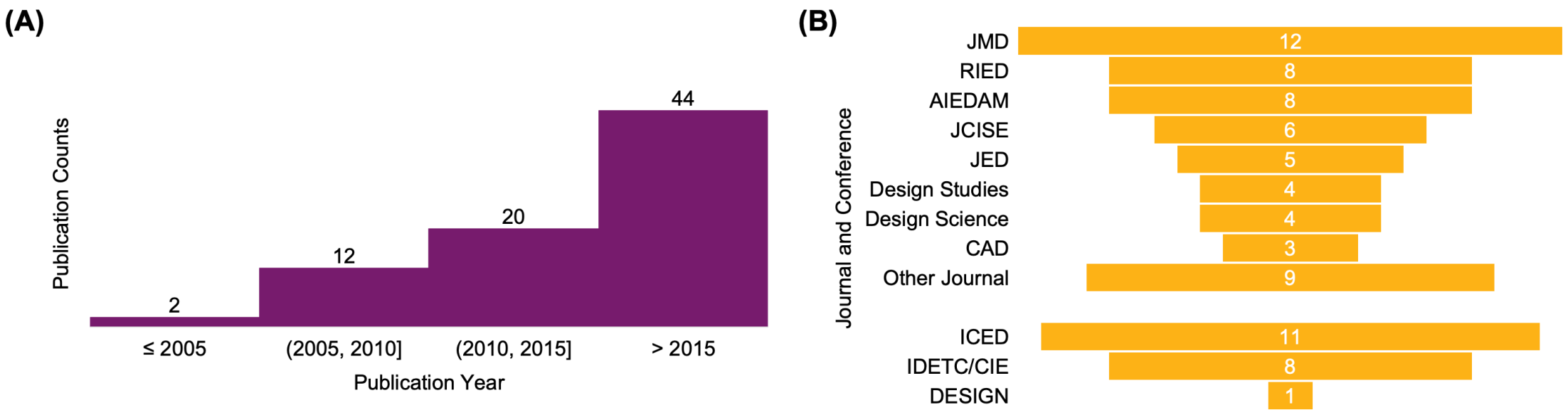

**Figure 1**: Distribution of the patent-for-design publications

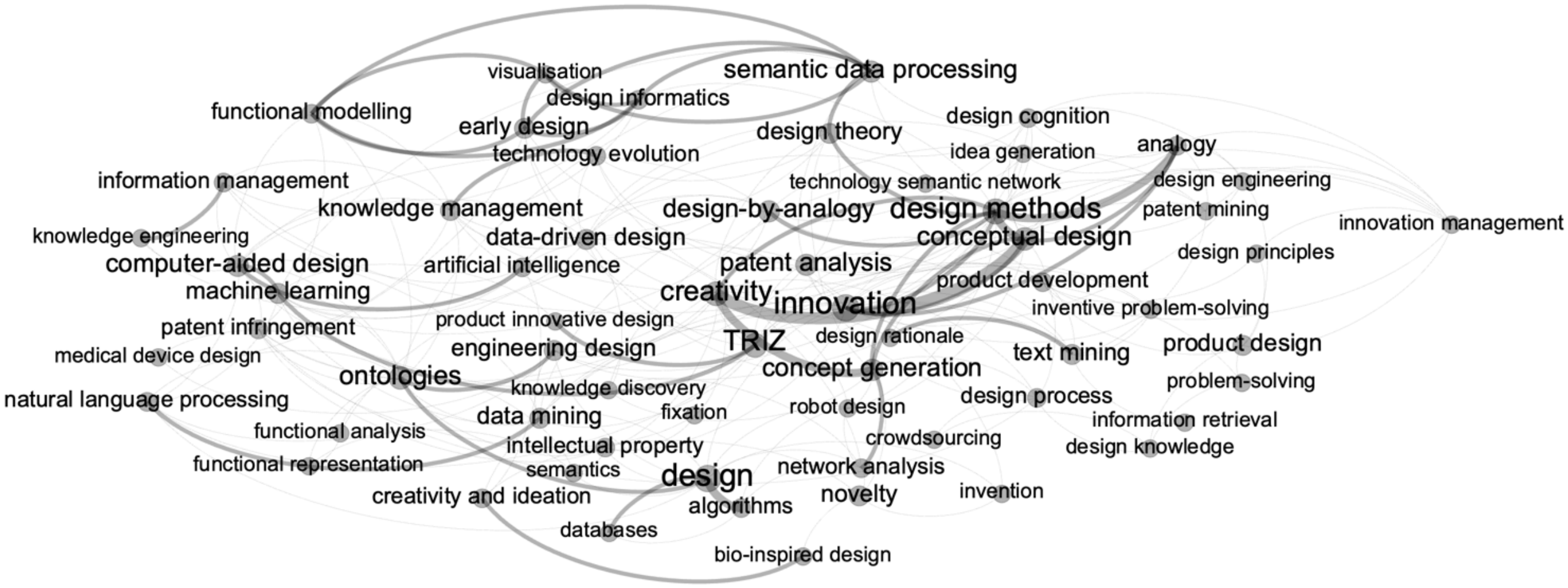

**Figure 2**: Co-occurrence of keywords of all retrieved literature

## 3 Patent Data

The patent documents are formally semi-structured. Patent documents contain bibliometric information (inventor, assignee, and publication time), classification information (IPC: International Patent Classification; USPC: US Patent Classification; CPC: Cooperative Patent Classification), citation information (patent and non-patent citations), text (title, abstract, claims, and detailed descriptions), and images (drawings and their descriptions). Fig. 3 shows an example of a patent document. In addition to the full document as a holistic representation of an invention, each specific part of the patent document has its own value in engineering design research. In the following section, we introduce how patent data are mined and analyzed.

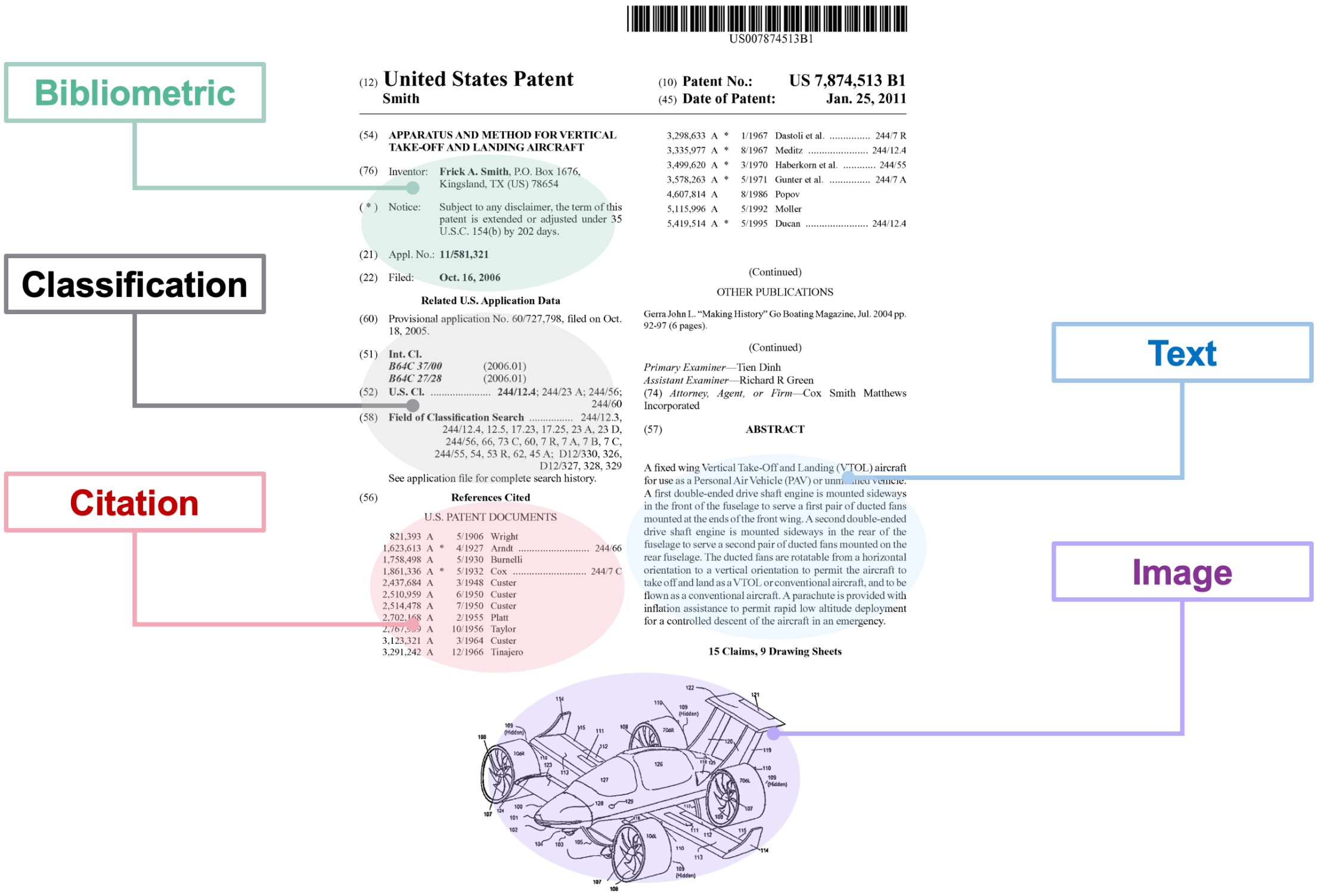

**Figure 3**: Example of a patent document (US Patent 7874513)

Table 2 summarizes the use of different information items of patent documents in engineering design literature. Almost half of these studies have used all fields of patent documents. Textual information received the most attention among all information items. Design information and knowledge were extracted from the patent titles, abstracts, and descriptions using various NLP techniques. Claims texts have also been mined to extract knowledge tuples for creating knowledge graphs [4,11,12]. Classification information, which contains categorical knowledge from patent examiners, can be viewed as annotations or labels for use in supervised machine learning tasks [5,13]. A few studies have also used classification information to cluster and manage existing design solutions and support design ideation [7,8,14].

**Table 2:** Parts of patent documents being used for engineering design research

| Parts of patent | Used in |
|---|---|
| Bibliometric information | [15–21] |
| Texts | Titles [6,9,22–24]<br>Abstracts [6,8–10,14,22–25]<br>Claims [4,8,10–12,23,26–28]<br>Descriptions [8,10,14,25,28–32]<br>Unspecified [17,33–52] |
| Images | [5,11,21] |
| Classification information | IPC [5,7,9,15,16,50,53–58]<br>CPC [8,10] |
| Citation information | [7,17,19,21,38,53–57] |
| Full document | [36,53,56,58–82] |

Table 3 lists several popular patent databases used by design researchers. The two leading patent offices (i.e., USPTO and EPO) provide researchers with various datasets of patent applications and granted patents. Researchers can directly download original US patent datasets from the USPTO Bulkdata website or PatentsView website in tab separated value (TSV) format for program readability. Both databases contain several patent documents and can be used in machine learning or deep learning pipelines that typically require big data for training and testing.

To promote data science research based on patent data, several communities have curated collections based on original patent documents and released some open tasks. NTCIR workshops, which were designed to support research in information access technologies [83], have been organizing patent-related campaigns and providing international patent collections since 2001. The topics of the campaigns included prior art search, patent classification, evaluation, and machine translation. Similarly, the CLEF-IP forum organized a series of shared tasks between 2009 and 2013, including patent retrieval, structure recognition, image classification and recognition, and novelty search, and provided the corresponding EPO datasets for benchmarking [84]. These curated patent collections can potentially support the development of relevant methods or tools to assist engineering design tasks such as design precedent search [17] and design novelty evaluation [57].

**Table 3**: Patent datasets

| Dataset | Source |
|---|---|
| USPTO | • https://bulkdata.uspto.gov<br>• https://patentsview.org/download/data-download-tables |
| EPO | • https://www.epo.org/searching-for-patents/data/bulk-data-sets.html |
| NTCIR | • http://research.nii.ac.jp/ntcir/data/data-en.html |
| CLEF-IP | • http://www.ifs.tuwien.ac.at/~clef-ip/download-central.shtml |

## 4 Engineering Design Research Based on Patent Data: A Detailed Review

During our review of patent-for-design literature, four themes emerged based on the contributions of individual publications: (1) design theory, (2) design methodology, (3) design tool, and (4) design strategy. These themes resonate with recent special issue topics in the data-driven engineering design journals [85–87]. Here, we provide a brief description of the four themes, which will serve as the basis for categorizing our literature review later in this paper.

***Design theory*** refers to the fundamental understanding of the design process and design rules as well as the causalities among designer traits, design preferences, behaviors, and performances. Design theory research often aims to identify the factors influencing design outcomes [53]. Spillers and Newsome [88] claimed that design theory is a meta-theory of existing theories of structures, machines, and aesthetics.

***Design methodology*** refers to a method or framework that can be repeatedly used to assist designers in carrying out design activities in one or multiple design phases, including: design need analysis, conceptual design, embodiment design, detailed or optimal design, and design evaluation [89]. Most design methodologies can be implemented manually or (semi-)automatically in computer-aided design tools.

***Design tool*** *r*efers to a tool that is developed based on one or more specific design methods to facilitate the design process. These tools typically have their own UI interfaces (web-based or software-based), such as TechNet [22] and InnoGPS[90]. In this study, we classified the works that the authors claimed to be practical tools in this category.

***Design strategy*** refers to a plan, policy, or process heuristic [91] that designers can use to guide design activities and processes [92]. For example, a design team can integrate the engineering design process with scientific research and entrepreneurship activities to achieve innovation [93]. The design strategy is macro, often motivated by the designers' own goals and intents, and in principle, need to be rationalized and guided by design theories.

Our literature analysis was organized into four categories [88]. Specifically, the first author of this paper categorizes these papers. The second author reviewed the categorization results in detail and proposed several changes. All authors discussed the categorization and reached a final agreement. Fig. 4 summarizes the literature distribution based on categories. It is noteworthy that alternative categorizations (e.g., different design phases [89]) can be used for different focuses and interests, whereas our categorization addresses the types of research contributions to design. In addition, one paper may contribute to one or more categories, as shown in Fig. 4B. Thus, the sum of the ratios of publications in each category over the total number of publications (78) is greater than 1, as shown in Fig. 4A.

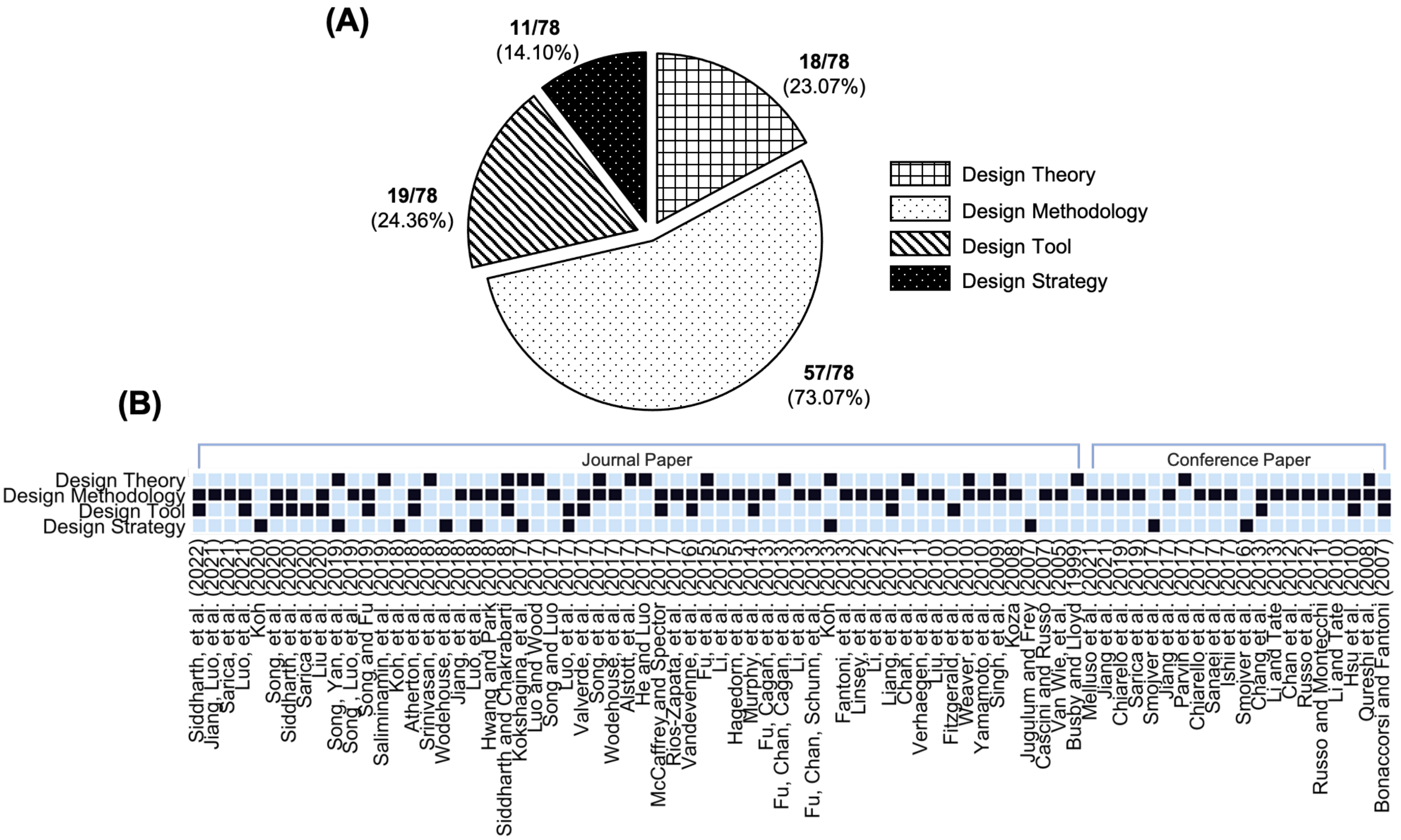

**Figure 4**: Distribution of patent-for-design publications by their research contribution types

### 4.1 Design Theory

Patent data serves as the empirical basis for research on various design theories. For example, Busby et al. [79] performed an experiment on the factors that influenced design solution search activities and used the patent database as the searching pool. Weaver et al. [75] and Singh et al. [76] developed transformation design principles by examining patents, products, and biological cases, assisting engineering designers in developing products by transforming or reconfiguring design knowledge between multiple states. Qureshi et al. [81] proposed a patent dissection method to discover product flexibility principles from a patent database.

Later, many researchers studied design-by-analogy (DbA) using patent databases as sources of design stimuli for drawing analogies [94]. They examined how analogical distance[34,36,53,56,73], commonness[73], and the modality of examples [73] influence DbA-based ideation outcomes. Similarly, Saliminamin et al. [61] used patents as idea triggers to explore the effect of precedents and design strategies on idea generation. Parvin et al. [82] used TRIZ contradiction models derived from specific patents to stimulate novel solutions for design problems.

In addition to the influencing factors on design outcomes, a few recent studies have mined bibliometric, citation, and classification information of patent documents to reveal the fundamental patterns in designers' exploratory behaviors[18], causality between design novelty and potential impact, [57] and growing complexity of the design and invention process[16]. Luo and Wood [16] used a set of patent-based metrics to discover the constantly increasing complexity of inventions and invention processes over time. According to their analysis inventions have been increasingly requiring larger and more diverse teams, as well as integrating a larger base of prior technologies to deliver more systemic and integrative new technologies. Alstott et al. [18] observed that most inventions are created by inventors who move between different technological domains during their careers, based on their analysis of 2.8 million inventors patenting records. They also identified that knowledge distance conditions inventors' abilities and performance in exploring new domains away from their prior domains. He and Luo [57] analyzed 3.9 million patents' citations as a proxy for prior art combinations and identified that the most valuable patented inventions are based on the integration of moderate mean novelty and high extreme novelty in the combination space. By analyzing non-patent literature in patent documents, they observed that the use of scientific and broader knowledge beyond patentable technologies increases the value of patented inventions.

### 4.2 Design Methodology

As shown in Fig. 4, most patent-for-design studies aim to develop new design methodologies. The earliest study relied on human expertise and manual efforts to extract design rules or heuristics from patents or use patents as design cases to illustrate design methods [19,34,36,53,59,61–63,65,66,68,71,73,79]. Recent studies have used NLP to identify innovative solutions from patent databases to design problems to facilitate the use of TRIZ [27,38,42,70]. For example, Cascini and Russo [42] proposed a method to automatically identify the design contradictions underlying a patented invention to support TRIZ. Li et al. [38] proposed an NLP-based methodological framework to classify patents based on invention level, as defined in TRIZ. Li et al. [27] presented an innovative design process that combines TRIZ and patent circumvention to develop new design solutions. On this basis, they further defined several rules for trimming technical features to avoid infringement during the design process and developed a TRIZ-based trimming method for patent design [70].

Another strand of research has developed design methodologies that leverage the design knowledge in patents for design representation and reasoning. For instance, Van Wie et al. [28] presented engineering products based on the function-behavior-structure (FBS) ontology [95] by analyzing patent documents. Bonaccorsi and Fantoni [47] expanded the functional basis by retrieving verbs from patents sampled from various IPC classes and classifying them to preserve the ontological function structure. They also developed a patent text analysis tool to demonstrate the usefulness of the expanded functional basis. Li and Tate [32] proposed a rule-based methodology to automatically retrieve functional requirements and design parameters from patent text using NLP techniques. Subsequently they introduced an enhanced study to represent a patent as a rule-based tree by combining NLP techniques and ontologies [52]. Jiang et al. [48] introduced a framework for building domain-specific ontologies based on a function analysis diagram to avoid patent infringement in the product development process. Yamamoto et al. [41] extracted subject-verb-object (SVO) tuples from patent texts for function division in conceptual design. Fantoni et al. [37] proposed an approach for extracting function–behavior–state information from patents. Liang et al. [39] and Liu et al. [40] proposed a methodology to discover design topics (or design rationales) from patent documents by combining text mining, clustering techniques, and term similarity metrics. Valverde et al. [67] developed a discovery matrix in which physical phenomena and technologies are matched by patents to inspire engineers. Hwang and Park [64] developed design heuristic sets for X (DHSfXs) from products and patents to facilitate concept design for specific goals. Atherton et al. [11] proposed a functional representation method using the annotation of geometric interactions derived from patent claims, which assists designers in better understanding prior designs. Chiarello et al. introduced NLP based methods to extract "advantages" and "drawbacks" from patent texts and a classification framework to organize the extracted knowledge [49], as well as introduced various methods to retrieve affordances from patent texts [44]. Melluso et al. [31] proposed an NLP-based methodology that leverages keyword search and rule-based entity matching techniques to discover patents that contain information about bad designs and design bias. Jiang et al. [45] introduced an NLP-based method to automatically extract design entities and the hierarchical and functional relations between them from patent text to identify the invention working principle.

Because patent databases are natural repositories of design precedents, methods of retrieving patents as inspirational stimuli to augment design ideation are of central interest to conceptual design researchers. For example, Verhaegen et al. [30] retrieved candidate patents as product precedents for analogical design by distilling product features. Rios-Zapata et al. [58] presented a creative design method that fuses combination and mutation models to support patent prior art search and analysis it in the early design stages. Russo and Montecchi [50] introduced a method that uses a prebuilt thesaurus of physical effects, WordNet for the abstraction of functions, and FBS ontology to search for prior art in a patent database. Similarly, Russo et al. [23] used the FBS ontology to build systematic queries to retrieve prior art patents for a given design problem by leveraging subject-object-action structures. Song et al. [56] proposed a method for patent stimuli searching based on community detection within a patent class network. Song and Luo [17] proposed a method for retrieving patent precedents for data-driven design by integrating searches through keywords, citations, and co-inventor networks. Jiang et al. [12] presented a framework that can assist designers in comparing prior arts and obtaining inspiration from them. Their framework is built on a functional knowledge graph with human-created ontologies. Liu et al. [9] proposed a method to extract functional terms from a patent database and processed them using clustering algorithms for design ideation. Luo et al. [7] used patent citation-based metrics to measure knowledge distance among different technology domains, and proposed workflows to search and retrieve design stimuli for analogy and synthesis across domains based on knowledge distance. Chan et al. [21] investigated whether visually similar design patents could be clustered together using a citation network of design patents.

Specifically, a group of studies has focused on patent data retrieval to support DbA. Linsey et al. [72] proposed the WordTree method to semantically re-represent design problems based on WordNet and guide designers to find potential analogies for innovative design from a set of patents. Fu et al. [14,25] used the combination of Bayesian model and latent semantic analysis (LSA) to map a set of patent documents in a network structure to guide patent searches for analogical inspirations. Murphy et al. [35] proposed a functional vector method based on the bag-of-words to encode patent documents into high-dimensional vectors for supporting an analogy search. Sanaei et al. [51] proposed an ensemble model to retrieve analogical solutions for a given problem in the entire patent database. Their ensemble model includes a word embedding model, graph matching model based on syntactic trees of sentences, transformation-based model that scores the similarity between sentences by measuring the partial transformations required, and long short-term memory (LSTM) model for deriving sentence embeddings.

The latest advances in data science and AI have enabled the development of automated or semi-automated design methods that process massive patent data. Koza [77] developed a genetic programming algorithm to solve design problems automatically and used a patent database to examine the novelty of the newly generated solutions. Wodehouse et al. [68] presented a clustering method for analyzing design opportunities using crowd intelligence. Song et al. [33] proposed a data-driven product platform design method based on core-periphery structure detection within functional word co-occurrence networks created from patent texts. Jiang et al. [5] proposed a convolutional neural network-based representation method for design images from patent documents to facilitate a visual DbA. Sarica et al. [6] presented an idea generation methodology based on a large technology semantic network of over four million technical terms using a word embedding model trained on a patent database [22].

### 4.3 Design Tool

Recent studies have leveraged patent data to develop data-driven design tools and facilitate and automate relevant design methodologies. Fitzgerald et al. [74] developed a design-for-environment (**DfE**) tool to manage and facilitate the analysis and reuse of successful products for conceptual design. Their DfE tool was a rule-based system built on TRIZ and DbA. Vandevenne et al. [29] developed a tool called scalable search for systematic biologically inspired design (**SEABIRD**), which enables designers to perform a scalable search for biological stimuli. SEABIRD uses rule-based text mining techniques to extract and map the product aspects of technical systems in patent documents and the organism aspects of biological systems in academic papers to identify candidate analogies. Chang et al. [46] introduced a methodology for design-around using TRIZ contradictions and text mining techniques. Hsu et al. [80] introduced a design process that represents patents with a design matrix inspired by axiomatic design theory, and uses matrix operations and TRIZ contradictions to develop innovative solutions by using a patent design-around process. They also developed a tool to assist designers develop innovative designs by using a design matrix. McCaffrey [96] developed **Analogy Finder**, a DbA support system, to identify adaptable semantic analogies from a patent database. Later, McCaffrey and Spector [69] devised a visual and verbal problem-solving representation to support human-machine collaboration in innovative design. Luo et al. [55] developed **InnoGPS,** a cloud-based tool that uses an empirically built interactive network map of all patent technology classes, to guide the search for design inspiration (from patent texts) and innovation opportunities in different domains. Using InnoGPS as the basis, Luo et al. proposed a series of data-driven design applications, including design opportunity identification [54,55] and analogical conceptual design [7,90]. Siddharth and Chakrabarti [65] developed **Idea-Inspire 4.0** and validated it on patents. Idea-Inspire represents both engineering concepts and biological ideas using SAPPhIRE model ontology [97] for biologically inspired design. Based on the Idea-Inspire tool and SAPPhIRE model, they developed an automated novelty evaluation method for engineering design solutions [60]. Song and Fu [8] used a topic modelling algorithm to structure a repository of mechanical design patents with three facets: behavior, material, and component. They developed a visual interaction tool for seeking design inspiration, named **VISION** [8]. Sarica et al. [22] applied word embedding techniques to patent data and constructed a large-scale technology semantic network of over four million terms called **TechNet**, which is accessible through API and a public web portal. TechNet has been used for design representation [98], prior art retrieval [24], idea generation [6], and concept evaluation [99]. Siddharth et al. [4] developed a large and scalable engineering knowledge graph based on the USPTO database, which can be used to support design inference, reasoning, and representation in engineering design applications.

### 4.4 Design Strategy

Patent databases also enable researchers to design strategies in both manual and automated ways. For example, Jugulum and Frey [78] studied a large number of inventions and summarized several general strategies used in these inventions as a taxonomy of concept design for improved robustness. Koh et al. [59,62,71] studied the proper methods and repercussions of reviewing patent documents during the early design stage. Kokshagina et al. [66] proposed the design-for-patentability' strategy to guide the innovation of engineering designers.

Several researchers have studied how to mine and analyze patent data to identify potential design directions and opportunities, and formulate long-term strategies for innovation at a higher level than the design process at an operational level. For example, Luo et al. proposed a series of patent-data-driven methods to enable macro-level design opportunity identification [54], strategic direction planning [7,55], and technology road mapping [15], using a total technology space map (TSM) based on the patent classification information. Smojver et al. introduced methodologies to develop longitudinal patent networks of specific design domains using bibliographic and citation data [19], and text mining [43] to support the visual analysis of technological change and evolution, which may inform macro-level design strategies.

## 5 Analysis of Methods and Algorithms

The patent-for-design literature has used a variety of research methods ranging from qualitative analysis and reasoning to the latest network analysis, data science, and machine learning techniques. Table 4 presents the categorization of papers according to the methods used.

**Table 4**: Methods and algorithms used in the patent-for-design publications

| **Methods** | **References** |
|---|---|
| Human-involved Study | [34,36,53,59,61–63,65,66,68,71,73,79,81] |
| Rule-based Expert System | [11,12,17,23,26–28,31,42,46,58,60,64,65,67,69,70,72,74–76,78,80,82] |
| Statistical Analysis | [14,16,18,20,57] |
| Genetic Programming | [77] |
| Machine Learning (Classification and Clustering) | [9,21] |
| Network Science | • Network visualization [7,8,10,14,15,19,33,43,54–56]<br>• Network-based metric analysis [7,15,33,43,56,57] |
| Vector Space Method | • Functional vector space [34,35]<br>• Technology space map [7,54,55]<br>• SAPPhIRE-based vector space [60]<br>• PA and OA space [29,30] |
| Deep Learning (Neural Networks) | • Artificial Neural Network [38]<br>• Convolutional Neural Network [5]<br>• Recurrent Neural Network [22,51] |
| Text Mining | • Text pre-processing [4,9,22–24,29,30,32,34,35,37–40,42–46,49–52,75,76]<br>• Syntactic analysis [4,32,37,41,44,45,49,51]<br>• Word co-occurrence analysis [33] |
| Topic Modelling | • Non-Negative Matrix Factorization [8,10]<br>• Latent Semantic Analysis [14,25,36] |
| Ontology, Semantic Network and Knowledge Graph | • Ontology [47,48]<br>• Semantic network [6,22,45,72]<br>• Automated or semi-automated constructed knowledge graph [4,37,39,40]<br>• Manually curated knowledge graph [11,12,26] |

Fig. 5 shows the co-uses of the methods and different information items of the patent data in the reviewed studies. Multiple methods can be used for various data sources. Each item was counted to ensure that each method matched the corresponding part of the patent data. The decomposition of classification information and text in Fig. 5 also refers to the summary in Table 2. Different line colors denote different parts of the patent data and the width of a line indicates the number of corresponding studies.

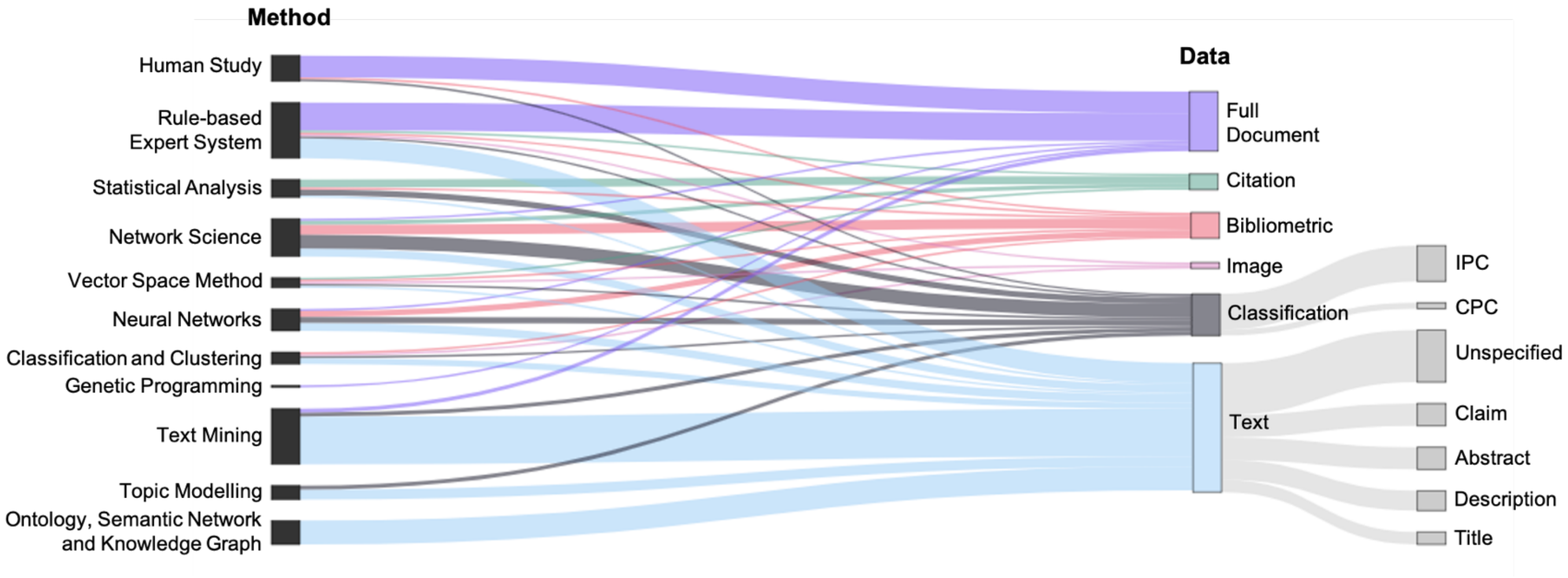

**Figure 5**: Methods in relation to different parts of patent data

Human subject experiments have primarily been used to develop theories and scientific understanding. In these studies, full patent documents were often directly provided to designers as design aids, such as prior art solutions [59,62,63,66,68,71,79,81] or design stimuli [34,36,53,61,65,73] for engineering design. In addition, a group of qualitative studies used knowledge-based rules or strategies to boost computer aided engineering design, and developed rule-based expert systems [11,12,17,23,26–28,31,42,46,58,60,64,65,67,69,70,72,74–76,78,80,82]. These expert systems typically require designers to collaborate with algorithms to address specific problems.

A wide range of data science techniques have been used in the patent-for-design literature. As shown in Fig. 6, patent-for-design studies using data science surpassed those without using data science in the past 10 years, and the gap is still expanding.[5] For instance, complex network analysis has been used to mine relational information from citations or measure proximities of semantic content among patents [7,8,10,14,15,19,33,43,54–56]. Using network-based metrics such as centrality, entropy, and coherence, prior studies have developed new scientific understandings of design artifacts and processes [15,57] and proposed patent data-driven design methodologies and strategies [7]. Some design tools, such as InnoGPS [7,55] and VISION [8], use network visualizations to guide designers in exploring design or technology spaces constructed on patent data or patent classification labels.

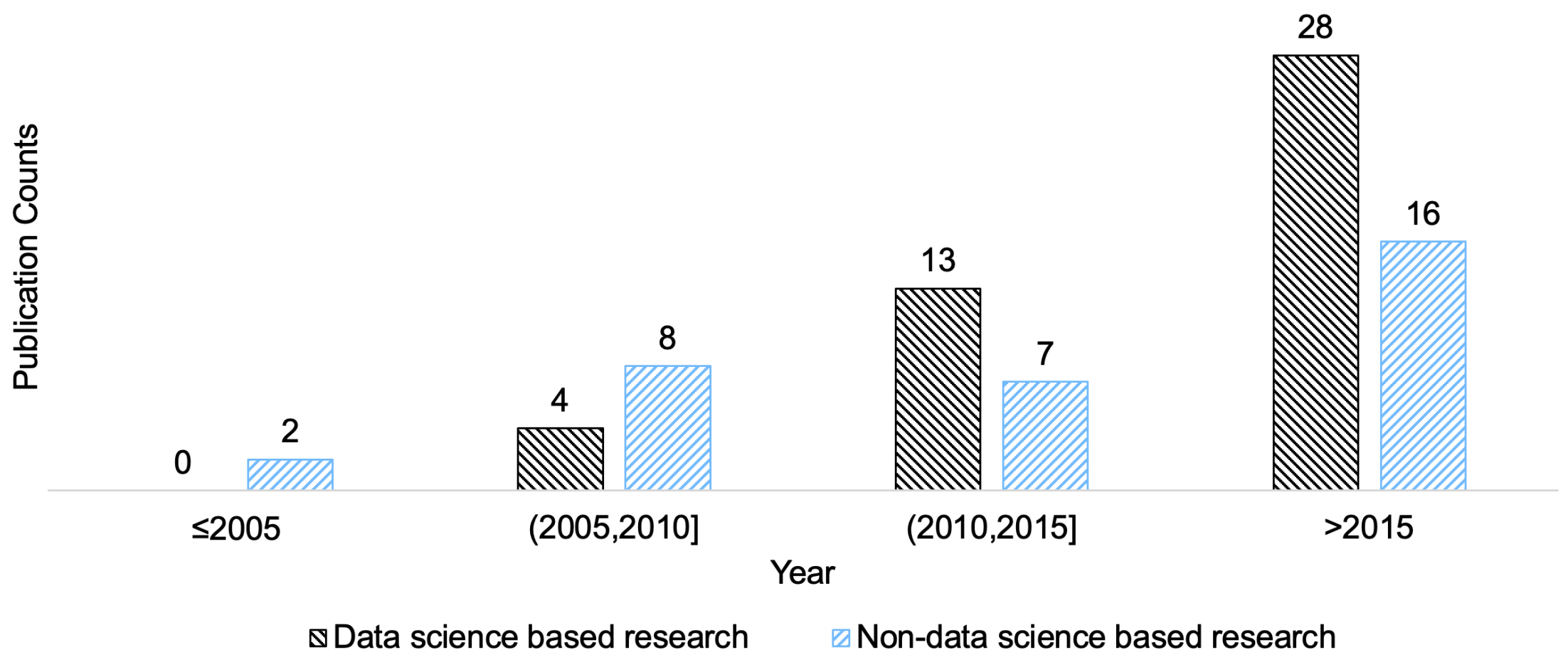

**Figure 6**: Growth of data science-based patent-for-design publications[5]

NLP techniques are increasingly being used to retrieve design knowledge from patent texts. Fig. 7 shows different types of semantic information extracted from the patent text, including specific semantic information (such as TRIZ contradictions), functional semantic information (such as functional verbs), and general semantic information (such as word embeddings).

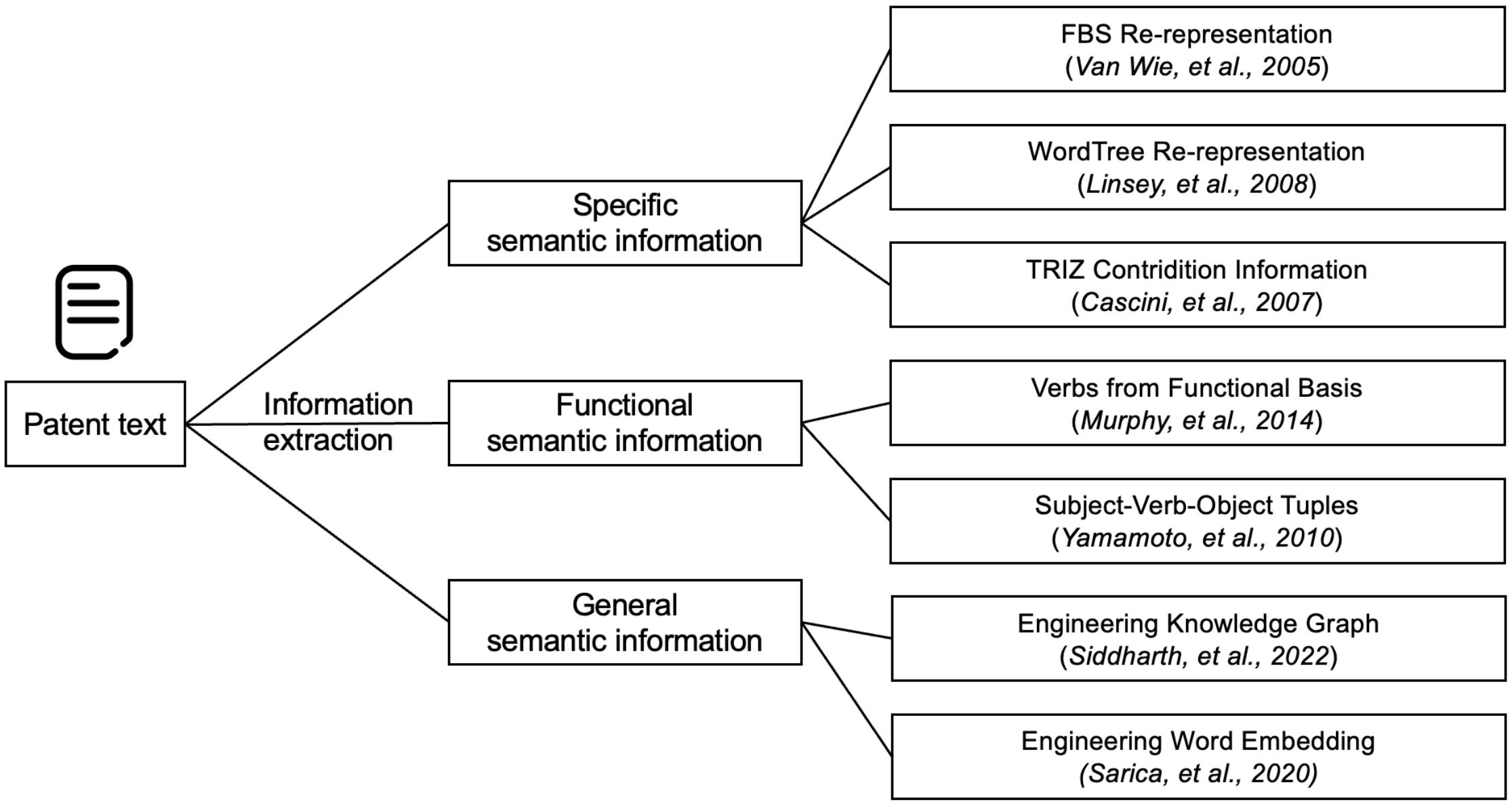

**Figure 7**: Information extraction from patent texts

[5] Figure 6 is manually developed based on the analysis and summary of methods and algorithms (Table 4). The data science-based patent-for-design publications stands for those that used advanced data mining and analysis techniques (including statistical analysis, network science, machine learning, deep learning, etc.) instead of directly using patent documents as design stimuli or case studies.

A few studies have used standard text pre-processing pipelines, including lemmatization, stemming, and stop-word removal techniques, for cleaning the raw text for further processing steps. In addition, design knowledge is extracted in certain forms of templates from a given patent text, using syntactic dependencies to support the automation of TRIZ [41] and the construction of knowledge graphs [4,37]. In addition, topic modelling algorithms, such as non-negative matrix factorization [8,10] and latent semantic analysis, [14,25,36] have been applied to patent texts to represent design repositories in a more structured form.

Semantic networks and knowledge graphs have long been used to support engineering design research [100]. Several early studies leveraged pre-trained common-sense semantic networks [72] or manually curated ontology-based knowledge graphs [11,12,26] to support design innovation and problem-solving. Until recently, the patent database was mined to construct large-scale cross-domain engineering semantic networks and knowledge graphs [4,22], serving as a knowledge infrastructure to support data-driven engineering design research and practice.

Deep neural networks, because of their ability to learn complex patterns from big data, have been used in patent-for-design literature. For instance, Li et al. [38] trained a neural network to classify patents based on the novelty level of an invention, as defined in TRIZ. Recently, researchers have developed design methodologies based on convolutional neural networks [5] and language models [22,51] trained on datasets of patent images and text, respectively. Other machine learning models, such as the naïve Bayes classification method [9], clustering methods[21], and advanced statistical analysis methods [14,16,18,20,57], have also presented their values in the patent-for-design literature. It is worth mentioning that all parts of patent documents have the potential to support engineering design research using diverse data science techniques, as shown in Fig. 5.

## 6 Future Opportunities and Directions

In this study, we reviewed and analyzed the patent-for-design literature to elucidate the status quo of this field. A clear increasing trend of using patent data in engineering design research (Fig. 2A) and a similar upward trend of using advanced data science techniques in the same literature (Fig. 6) can be observed. Although existing studies have already shown the value of patent databases for data-driven engineering design research and practice, there are still several challenges and opportunities regarding 1) patent data, 2) data science algorithms, and 3) design applications. In the following section, we discuss research opportunities and map the course of feasible directions for future studies.

### 6.1 Research Opportunities Regarding Data

Patent databases are natural benchmark datasets for supervised machine learning applications because every patent is rigorously labelled by patent offices regarding its technological domain(s). The classification information and citation-based metrics of patents can serve as the gold standard that enables training, benchmarking, testing, and comparison of the performance of different supervised learning algorithms [13].

Patent texts contain rich design information that can be used to create datasets for NLP-related tasks such as entity recognition and detailed analysis of functions, behaviors, and structures of engineering designs. However, these datasets require heavy labelling and annotation to curate, and thus, are laborious human tasks requiring considerable expertise, time, and resources.

In Fig. 5, we can see that bibliometrics, images, citations, and classification information of patents are not mined as commonly as textual information for design support. It is recommended that researchers use multimodal patent information instead of a single modality to develop a more systemic understanding of design artifacts and processes or more intelligent design methods and tools.

Furthermore, the current patent-for-design literature focuses on USPTO and EPO patents as data sources (primarily because they are written in English). We believe that patents from other countries such as Japan[6] and China[7] are also useful for supporting engineering design research.

### 6.2 Research Opportunities Regarding Algorithms

Owing to their large volume, patent databases are suitable for implementing modern deep learning techniques for specific tasks that typically require massive amounts of data for model training. Rapidly advancing data science technologies have also empowered the research community to work on patent databases, and many statistical, graph theoretical, and deep learning methods have been developed or adopted for various tasks and produced state-of-the-art results. Some of these studies may shed light on new research opportunities for engineering design researchers to better use patent data. For instance, retrieving prior art or relevant patents to derive design inspiration or to prevent patent infringement is an important step in product development. Engineering design researchers may adopt or adapt deep learning methods developed in computer science literature for patent prior art search [101,102], classification [103–105], and representation [103,106–108].

[6] China National Intellectual Property Administration: https://english.cnipa.gov.cn/
[7] Japan Patent Office: https://www.jpo.go.jp/e/

Specifically, the latest deep learning capabilities for natural language understanding (NLU) have great potential for enhancing design-knowledge representations. Large language models, such as bidirectional encoder representations from transformers, if trained on massive patent data, can enable high-dimensional statistics to derive accurate semantic relations and meanings of engineering concepts, and support the creation of comprehensive design knowledge bases [109,110]. In turn, such semantic knowledge bases may offer not only structural and explicit information in patent texts but also latent causal relations and working mechanisms in technical inventions, and complement the existing common sense knowledge bases for use in design [100].

In addition, recent progress in graph neural networks (GNN) has enabled us to extract relational information among patents and derive high-dimensional representations for downstream tasks [111], such as design repository restructuring and design stimuli identification. Provided that patent documents normally contain multimodal information on design, we can take advantage of multimodal deep learning techniques to develop more systematic and informative design representations[112], such as multimodal knowledge graphs and resultant graph embeddings. The multimodal deep learning of patent data has tremendous potential for engineering design applications.

Finally, in the past five years the AI community has seen rapid development of powerful deep generative models (e.g., VAE [113] and GAN [114]) and large pre-trained language models (e.g., GPT-3 [115] and BERT [116]) for image and text generation. These generative models can also be used to develop AI-aided creative ideation or design-generation methods. Researchers can develop generative models [117] specifically for design synthesis and analogy by learning engineering design-related knowledge from the patent data. For instance, Zhu and Luo [118] recently used patent text data in different design domains to fine-tune the OpenAI GPT-2 model to train virtual domain experts, and then used them to generate novel design concepts with knowledge from specific target domains.

### 6.3 Research Opportunities Regarding Applications

Fig. 4 shows that patent data are mostly used to develop design methods and tools, whereas applications in design theory and strategy research at higher levels deserve further exploration. Large-scale patent databases that contain detailed multimodal content and rich bibliometrics, citations, and classification information offer unprecedented opportunities for design theory building. For example, big patent data analysis, in contrast to small-sample human subject studies, may offer statistically significant findings that explain the behaviors and performances of design agents across diverse technological domains [18] and the conditions for achieving highly valuable designs and breakthrough innovations [57]. However, only a few studies have used big data-driven methods to build theories in the field of design science.

Moreover, although the literature has reported several design tools developed based on corresponding design methodologies, only a few are open-sourced and publicly accessible, and can be directly used by other engineering designers or researchers. To generate greater real-world impact and foster improvements of such tools, we recommend researchers to open source their developed tools (such as uploading the codes on GitHub) and provide open tool access to the public to use, test, and provide feedback.

In addition, we believe that the statistics of the big multimodal patent database could substantially deepen our fundamental understanding of design teamwork, designer behaviors and rationales, design impact dynamics, etc. Existing patent-for-design studies have demonstrated the potential of mining patent databases to explore white space and identify feasible directions for the R&D activities of designers, design teams, and large companies [55]. Several patent data providers, such as Patsnap[8] and

[8] Patsnap Pte. Ltd.: https://www.patsnap.com/

Incopat[9], have begun to provide patent analytics to inform innovation strategies and management decisions. Thus, we recommend that researchers explore and experiment with various AI methods on billions of patent documents and unlock the potential of data-driven design strategy, management, and innovation decisions [119].

### 6.4 Inspiration from Patent Analytics in Other Fields

This study focuses on engineering design research. Meanwhile, research in other fields also mined and analyzed patent data from different perspectives and for other interests beyond those in the engineering design community. For example, a global search in the Web of Science database using the keyword "patent data" would also return publications from journals such as *Management Science*, *Research Policy*, *Scientometrics*, and *World Patent Information* which use patent data as the empirical basis. These studies may also provide inspiration for design research opportunities.

For instance, management researchers have mined patent data to inform financing and capital decisions [120] and analyze market dynamics [121] by correlating the patenting behaviors of companies and businesses, as well as the economy. Some studies have aimed to establish the relationship between a company's patent records and its future growth and success [122]. Analogically, engineering design researchers may also analyze designers' patent records to infer their innovation-related behaviors and performances, and predict their career prospects [18].

[9] Incopat Co. Ltd.: https://www.incopat.com/

Patent value estimation and prediction are of interest in scientometric and econometric research. For example, Bakker [123] identified a linear relationship between patent citations and patent value. Bass and Kurgan [124] used machine learning techniques to identify the key factors influencing patent value, including the past performance of inventors, assignees, and the number of referenced patents. Du et al. [125] developed a recommendation system for high-quality patent trading based on the hybrid analytics of patent texts, classifications, and citations. Similarly, data-driven design evaluation (employing patents as a proxy for designs or evaluation benchmarks) is also of central interest to engineering design researchers [57].

Furthermore, several researchers have leveraged patent data to characterize and classify product designs. For instance, Chan et al. [126] combined clustering methods with experimental validation to identify product styles from over 350,000 US design patents. Huenteler et al. [127] extracted the hierarchy level in a product's design (system, subsystem, and component) from patent citation networks. Patents have also been studied as proxies for intellectual property. For instance, Lee and Hsiang [128] fine-tuned the OpenAI GPT-2 model to automatically generate patent claim text for patent filing. Inspired by these studies, engineering researchers may mine patent data to gauge trends in design innovation [16] and train generative neural networks to generate novel designs for innovation [118].

## 7 Conclusion

Patent databases are ideal resources for engineering design researchers to develop design theories, strategies, methods, and tools based on manual efforts and various data science techniques because of the richness of the design information contained in patent documents. The past decade has witnessed a growing trend in the use of data science techniques to mine and analyze patent databases to support engineering design research and practice. This study contributes to the patent-for-design literature by elucidating the status quo of this field and identifying the potential research opportunities and directions. We believe that our review and propositions can serve as a guide for design researchers and practitioners in discovering the greater value of patent databases for engineering design research, developing more powerful and intelligent patent data-driven design methods and tools, and advancing our scientific understanding and theories of design.

There are several limitations in using patent data to support the engineering design process. First, some patents fields for IP strategies by companies (such as patent fence strategy) might not represent actual engineering design outcomes and lead to bias when researchers count them in the statistics. In addition, the correctness of the detailed information in patent documents may not be checked by examiners who primarily focus their analysis on the patentability, novelty, and inventive steps of patent applications. For comparison, the validity of the technical information in published scientific papers will be checked by peer reviewers. Although this review paper has only focused on leveraging patent data for engineering design research and practice, the AI and data science techniques that we analyzed can be used to mine and analyze a wider range of data for data-driven design and innovation [127], such as scientific papers, reports, books, products, and CAD databases[129].

Owing to the richness of patent data, researchers from various fields have used patent data to acquire indicators related to economics, technology, and innovation management [130–133]. In addition, the vast amount of technical knowledge accumulated in patents has created a basis for automating parts of the design process or the discovery of new materials and drugs [134,135]. Researcher may find inspiration from these studies in other fields and leverage their insights into engineering design. In addition, several recent startups and companies that focus on AI- and data-driven innovative designs have attracted much attention from venture capitalists, such as Small Design[10] (as of Jan 01, 2022, it raised $15M in total) and Tezign[11] (as of Jan 01, 2022, it raised $150M in total). In this case, the commercial value of mining patent databases for automated engineering design may exist.

In conclusion, given the continual accumulation of patent data and rapid advancements in data science, machine learning, and artificial intelligence capabilities, the value and importance of patent data-driven approaches to engineering design and innovation are expected to grow in the future. Our review addresses this trend and aims to stimulate and guide future efforts to exploit and explore opportunities for patent data-driven engineering design and make the design process more intelligent.

[10] https://www.smalld.cn/
[11] https://www.tezign.com/

**Acknowledgement**

The authors acknowledge the funding support for this work received from the SUTD Data-Driven Innovation Laboratory (DDI, https://ddi.sutd.edu.sg/), National Natural Science Foundation of China [52035007, 51975360], Special Program for Innovation Method of the Ministry of Science and Technology, China [2018IM020100], National Social Science Foundation of China [17ZDA020].